# Multi-photon boson-sampling machines beating early classical computers

Hui Wang[1,2,*], Yu He[1,2,*], Yu-Huai Li[1,2,*], Zu-En Su[1,2], Bo Li[1,2], He-Liang Huang[1,2], Xing Ding[1,2], Ming-Cheng Chen[1,2], Chang Liu[1,2], Jian Qin[1,2], Jin-Peng Li[1,2], Yu-Ming He[1,2,3], Christian Schneider[3], Martin Kamp[3], Cheng-Zhi Peng[1,2], Sven Höfling[1,3,4], Chao-Yang Lu[1,2,$], and Jian-Wei Pan[1,2,#]

[1] Shanghai Branch, National Laboratory for Physical Sciences at Microscale and Department of Modern Physics, University of Science and Technology of China, Shanghai, 201315, China

[2] CAS-Alibaba Quantum Computing Laboratory, CAS Centre for Excellence in Quantum Information and Quantum Physics, University of Science and Technology of China, China

[3] Technische Physik, Physikalisches Instität and Wilhelm Conrad Röntgen-Center for Complex Material Systems, Universitat Würzburg, Am Hubland, D-97074 Würzburg, Germany

[4] SUPA, School of Physics and Astronomy, University of St. Andrews, St. Andrews KY16 9SS, United Kingdom

* These authors contributed equally to this work
$ cylu@ustc.edu.cn, # pan@ustc.edu.cn

**Boson sampling is considered as a strong candidate to demonstrate the "quantum computational supremacy" over classical computers. However, previous proof-of-principle experiments suffered from small photon number and low sampling rates owing to the inefficiencies of the single-photon sources and multi-port optical interferometers. Here, we develop two central components for high-performance boson sampling: robust multi-photon interferometers with 99% transmission rate, and actively demultiplexed single-photon sources from a quantum-dot-micropillar with simultaneously high efficiency, purity and indistinguishability. We implement and validate 3-, 4-, and 5-photon boson sampling, and achieve sampling rates of 4.96 kHz, 151 Hz, and 4 Hz, respectively, which are over 24,000 times faster than the previous experiments, and over 220 times faster than obtaining one sample through calculating the matrices permanent using the first electronic computer (ENIAC) and transistorized computer (TRADIC) in the human history. Our architecture is feasible to be scaled up to larger number of photons and with higher rate to race against classical computers, and might provide experimental evidence against the *Extended Church-Turing Thesis*.**

Quantum computers[1] can in principle solve certain problems faster than classical computers. Despite substantial progress in the past two decades[2-4], building quantum machines that can actually outperform classical computers for some specific tasks—an important milestone termed as "quantum supremacy"—remained challenging. In the quest of demonstrating the "quantum supremacy", boson sampling, an intermediate (i.e., non-universal) quantum computer model proposed by Aaronson and Arkhipov[5], has received considerable interest as it requires much less physical resources than building universal optical quantum computers[6].

A quantum boson-sampling machine can be realized by sending $n$ indistinguishable single photons through a passive $m$-mode ($m > n$) interferometer, and sampling from the probabilistic output distribution. Mathematically, the probability amplitude of each output outcome is proportional to the permanent of a corresponding $n \times n$ submatrix, which is strongly believed to be intractable because calculating the permanent is a so-called #P-complete complexity problem. Note that, however, boson sampling is itself not a #P-complete problem, i.e., cannot efficiently calculate the matrix permanent. For a specifically defined task of sampling over the entire distribution, it is expected that a sufficiently large quantum boson-sampling machine cannot be efficiently simulated by the classical computers[5,7,8]. In principle, a large-scale boson-sampling machine would constitute an effective disproof against a foundational tenet in computer science: the *Extended Church-Turing Thesis*, which postulates that all realistic physical systems can be efficiently simulated with a (classical) probabilistic Turing machine.

To this end, an experimental roadmap for demonstrating "quantum supremacy" is to construct multi-photon boson-sampling machines with increasing number of input photons and faster sampling rates to race against classical computers. However, the overall performance of the previous proof-of-principle boson-sampling experiments[9-17] were critically limited due to the lack of high-quality single-photon sources and low-loss multi-mode circuits. For example, the most commonly used pseudo-single photons created using spontaneous parametric down-conversion[18] (SPDC) were intrinsically probabilistic and mixed with multi-photon components. The SPDC probability was kept small (about a few percent) in order to suppress the unwanted two-photon emission. The frequency correlation of the SPDC photon pairs and the inefficient collection into single-mode fibers further reduced the single-photon heralding efficiency to typically a low level of ~1% in the previous work[9-16] (see Supplementary Information Table S1). In addition, the boson-sampling rate was significantly reduced due to the coupling and propagation loss in the multi-mode photonic circuits. In an attempt to solve the intrinsic probabilistic problem of SPDC, spatial or temporal multiplexing[19,20] and scattershot boson sampling[21] schemes were proposed and demonstrated[14]. Yet, so far, all the previous quantum optical boson-sampling machines[9-17] have demonstrated only up to three single photons with arbitrary input configurations and 4-6 photons in special Fock states, and the obtained sampling rates were several orders of magnitudes too low to even outperform some of the earliest classical computers.

**Indistinguishable single photons**

Scaling up boson-sampling to large number of photons and with high sampling rates represents a non-trivial experimental challenge. Importantly, it requires high-performance single quantum emitters[22-24] that can deterministically produce one and only one photon under each pulsed excitation. The generated photons must simultaneously have high single-photon purity (that is, the multi-photon probability should be vanishingly small), high indistinguishability (that is, photons are quantum mechanically identical to each other), and high collection efficiency into a single spatial mode[25-27]. These three key features are compatibly combined in our experiment using pulsed *s*-shell resonant excitation[28] of a single self-assembled InAs/GaAs quantum dot embedded inside a micropillar cavity[29-31] (see Fig.1 and Supplementary Information).

At π pulse excitation with a repetition rate of 76 MHz, the quantum dot-micropillar emits ~25.6 million polarized, resonance fluorescence single photons per second at the output of a single-mode fiber, of which ~6.5 million are eventually detected on a silicon single-photon detector. Considering the detector dead time of ~42 ns, the actual count rate should be corrected to 9 MHz (Fig. 2a). This is the brightest single-photon source reported in all physical systems to date, which are directly used—without any spectral

filtering—for the photon correlation and interference measurements, and for boson sampling. We measure its second-order correlation, and observed $g^2(0) = 0.027(1)$ at zero time delay, which confirmed the high purity of the single-photon Fock state. We perform Hong-Ou-Mandel interference as a function of the emission time separation between two single photons[31]. With a time separation of 13 ns and 14.7 μs, photon indistinguishabilities of 0.939(3) and 0.900(3) are measured, respectively (see Fig. 2b and Supplementary Information). Thanks to the pulsed resonant excitation method that eliminates dephasings and time jitter[28], we obtain long streams near-transform-limited single photons that are sufficient for multi-photon experiments on a semiconductor chip for the first time.

**Efficient multi-photon source**

Next, we de-multiplex the single-photon stream into different spatial modes using fast optical switches that consist of Pockels cells (with a transmission rate >99% and extinction ratio >100:1) and polarizing beam splitters (with an extinction ratio >1200:1). The Pockels cells, synchronized to the pulsed laser and operated at 0.76 MHz with a rising time of 8 ns, convert the single-photon pulse train into 3, 4, or 5 separate beams (see Supplementary Information and Fig. S5). The largest time separation between two de-multiplexed photons is ~1.05 μs (80 pulses), where the photon indistinguishability remains 0.923 (Fig. 2b).

To ensure that these pulses arrive simultaneously at a multi-mode interferometer, optical fibers of different lengths and translation stage are used to finely adjust their arrival time. The average efficiency of the optical switches is ~84.5%, which was mainly due to the coupling efficiency and propagation loss in the optical fibers. The efficiency can be improved in the future using faster Pockels cells (see Supplementary Information). Thus, we eventually obtain five separate single-photon sources with end-user efficiencies of about 28.4%. Note the active de-multiplexing method eliminates the common technical overhead for overcoming the inhomogeneity of independent self-assembled quantum dots to build many identical sources.

**Ultra-low-loss photonic circuit**

Another important ingredient for reliable and fast boson-sampling is a multi-mode interferometric linear optical network that is phase stable, has high transmission rate, and can implement a Haar-random unitary matrix. While the previously demonstrated waveguide-based photonic chips showed promise for large-scale integration[10-16], the coupling and propagation loss in these chips seriously limited the overall efficiencies to ~30% so far (see Supplementary Information Table S1).

Here, we put forward a new circuit design that simultaneously combines the stability, matrix randomness, and ultra-low transmission loss. As shown in Fig. 1 (see also Fig. S6), a 9×9 mode interferometer is constructed with a bottom-up approach, from individual tiny trapezoid, each optically coated with polarization-dependent beam splitting ratios (Supplementary Information). This network consists of 36 beam splitters and 9 mirrors, and implements a near-unitary transformation to input state (Fig. 2c, d). Thanks to the antireflection coating, the overall transmission efficiency (from input to output) is measured to be above 99%. By Mach-Zehnder-type coherence measurements, the spatial-mode overlap is determined to better than 99.9%. The interferometer is housed on a temperature-stabilized baseplate, and remains stable at least for weeks (for a test, see Fig. S7). Such a design can be further improved[32] and scaled up to reasonably larger dimensions, which can be sufficient for the near-term goal of demonstrating quantum supremacy through boson sampling.

**Experimental results and validation**

We send three, four, and five single photons into the 9-mode interferometer, and measure the output multi-photon events, as shown in Fig. 3. We use nine silicon single-photon avalanche detectors (efficiency ~32%), one in each output of the interferometer, to register the no-collision (one photon per output-mode) events, which have 84, 126, and 126 different output distributions for the 3-, 4-, and 5-boson sampling, respectively. A total of 446084 three-photon events (Fig. 3a), 36261 four-photon events (Fig. 3b), and 11660 five-photon events (Fig. 3c) are obtained in accumulation time of 90s, 240s, and 2900s, respectively. The obtained data (solid bar, denoted as $q_i$) are plotted together with ideal probability distribution (empty bar, denoted as $p_i$) in Fig. 3. We quantify the match between these two sets of distributions using the measure of similarity, defined as $F = \sum_i \sqrt{p_i q_i}$, and the measure of distance, defined as $D = (1/2)\sum_i |p_i - q_i|$. From the data in Fig. 3, we can calculate similarities of 0.984(1), 0.979(5), and 0.973(9), and distancess of 0.125(1), 0.141(3), and 0.178(5) for the 3-, 4-, and 5-boson sampling, respectively.

For a large-scale boson-sampling device, not only the calculation of its outcome, but also a full certification of the outcome is strongly conjectured to be intractable for classical computation. There have been proposals[33-35] and demonstrations[15,16] for validating boson-sampling that can provide supporting or circumstantial evidence for the correct operation of this protocol. In our work, we first employ Bayesian analysis[34] to rule out uniform distribution (Fig. 4a). With only ~20 events, we can reach a confidence level of 99.8% that these outcomes are from genuine boson-samplers. Another possible hypothesis is using distinguishable single photons (classical particles) or spatial-mode mismatched interferometers, which should be excluded by applying standard likelihood ratio test[35]. Figure 4b shows an increasing difference between solid (indistinguishable bosons) and dotted lines (distinguishable bosons) as experimental events increasing, and thus the distinguishable hypothesis is ruled out with only ~50 events (see Supplementary Information).

**Conclusion and outlook**

Owing to our development of the high-efficiency source of highly indistinguishable single photons and ultra-low-loss photonic circuits, the experiment demonstrated 3-boson sampling rate of 4.96 KHz is ~27,000 times faster than the best previous experiments using SPDC[9-16], and ~24,000 times faster than the recent work[17] using passive demultiplexing (thus intrinsically inefficient) of quantum-dot single photons using incoherent excitation that limited the photon indistinguishability to 52%-64%. Meanwhile, we achieve the first 4- and 5-boson sampling using single-photon Fock

state—which were formidable challenges before—and obtain high sampling rates of 151 Hz and 4 Hz, respectively. These multi-photon boson-sampling machines have also reached a computational complexity that can race against early classical computers. Under the specific racing rule in ref. 5, 9, 10, we could compare the required time for obtaining one output sample using the quantum machines with the simulated time for calculating one permanent using the published data of the early classical computers (see Supplementary Information). As shown in Table SII, the quantum photonic machines are provably faster for the boson-sampling task than ENIAC and TRADIC, the first electronic computer and transistorized computer.

Our work has demonstrated a clear, realistic pathway to build boson-sampling machines with many photons and fast rates. Using superconducting nanowire single-photon detectors[36,37] with reported efficiency of ~95% and antireflection optical coating, one can straightforwardly increase the 3-, 4-, and 5-boson sampling rates to 130 KHz, 12 KHz, and 1 KHz, respectively, and implement 14-boson-sampling with a count rate of 5/h (see Supplementary Information). A remaining challenge is to remove the cross-polarization in the confocal setup—used to extinguish the laser background—which reduced the single-photon source efficiency by half. Future work will focus on deterministic dot-micropillar coupling[38] and developing side excitation[39] to boost the single-photon source efficiency to over 74%, in which case we can expect 20-boson sampling rate of ~130/h, and an increasing quantum advantage over classical computation for larger number of photons.

**Acknowledgements:**

We thank S. Aaronson, B. Sanders, and P. Rohde for helpful discussions. This work was supported by the National Natural Science Foundation of China, the Chinese Academy of Sciences, the National Fundamental Research Program, and the State of Bavaria.

**Author contributions:**

C.-Y.L. and J.-W.P. conceived and designed the experiment, C.S., M.K., and S.H. grew and fabricated the quantum dot samples. H.W., Y.H., Y.-H.L., Z.-E.S., B.L., H.-L.H., X.D., M.-C.C., C.L., J.Q., J.-P. L., Y.-M.H., C.S., M.K., C.-Z.P., S.H., and C.-Y.L. performed the experiment, S.H., C.-Y. L., and J.-W.P. analyzed the experimental data, and C.-Y.L. and J.-W.P. wrote the paper.

**Author Information:**

The authors declare no competing financial interests. Correspondence and requests for materials should be addressed to C.-Y.L. (cylu@ustc.edu.cn) or J.-W.P. (pan@ustc.edu.cn).

**Data availability:**

The data that support the plots within this paper and other findings of this study are available from the corresponding author upon reasonable request.

**Figure captions:**

**Figure 1 | Experimental setup for multi-photon boson-sampling.** The setup includes four key parts: the single-photon device, de-multiplexers, ultra-low-loss photonic circuit, and detection. The single-photon device is a single InAs/GaAs quantum dot coupled to a 2-μm diameter micropillar cavity, which yields a Purcell factor of 7.63(23) at resonance. The quantum dot is coherently pumped by a picosecond laser. A confocal microscope is operated in a cross-polarization configuration to extinguish laser background. The resonance fluorescence single photons collected into a single-mode fiber are sent to active de-multiplexers, which consist of Pockels cells and polarizing beam splitters, and separated into five spatial modes. The five photons are then fed into a tailor-made ultra-low-loss photonic circuit that consists of 36 beam splitters. Finally, the output out of the interferometer are measured by nine single-photon detectors and the multi-photon coincidence are analyzed by a time-to-digit converter (TDC).

**Figure 2 | The single photon source and interferometer for boson-sampling. a**,

Observed Rabi oscillation by pulsed resonant excitation of the quantum dot. The blue dots are directly measured by silicon detectors, whereas the red dots are corrected by the dead time of the detectors. The single-photon counts reach maximum at the π pulse power, which is 1.6 nW. **b**, The measured photon indistinguishability drops slightly from 0.939(3) at 13 ns to a plateau of 0.900(3) at >10 μs separation, fitted with a decaying time constant of 2.1 μs, assuming non-Markovian noise model. The blue arrow indicates the regime in our current work where two photons are maximally separated by a time of 1.05 μs due to de-multiplexing. The error bars denote one standard deviations, deduced from propagated Poissonian counting statistics of the raw photon detection events. **c, d**, Measured elements (**c**, amplitude and **d**, phase) of the unitary transformation of the optical network.

**Figure 3 | Experimental results for the (a) 3-, (b) 4-, and (c) 5-boson sampling.** The measured relative frequencies of all no-collision output combinations, denoted by ($i, j, \cdots$) where there is one photon detected in each output mode $i, j, \cdots$. The solid bars are the normalized coincidence rate of different output distribution. The empty bars are theoretical calculations in the ideal case. The error bar is one standard deviation from Poissonian counting statistics.

**Figure 4 | Validating boson sampling results.** The open points in **a** and the dotted lines in **b** are tests applied on simulated data generated from the two alternative hypotheses, sampling from a uniform distribution and distinguishable particles, respectively. In both **a** and **b**, the solid points and solid lines are tests applied on the experimental data. A counter is updated for every event and a positive value validates the data being obtained from a genuine boson sampler. **a**, Application of the Bayesian analysis to test against uniform distribution. **b**, Discrimination of the data from a distinguishable sampler using standard likelihood ratio test.

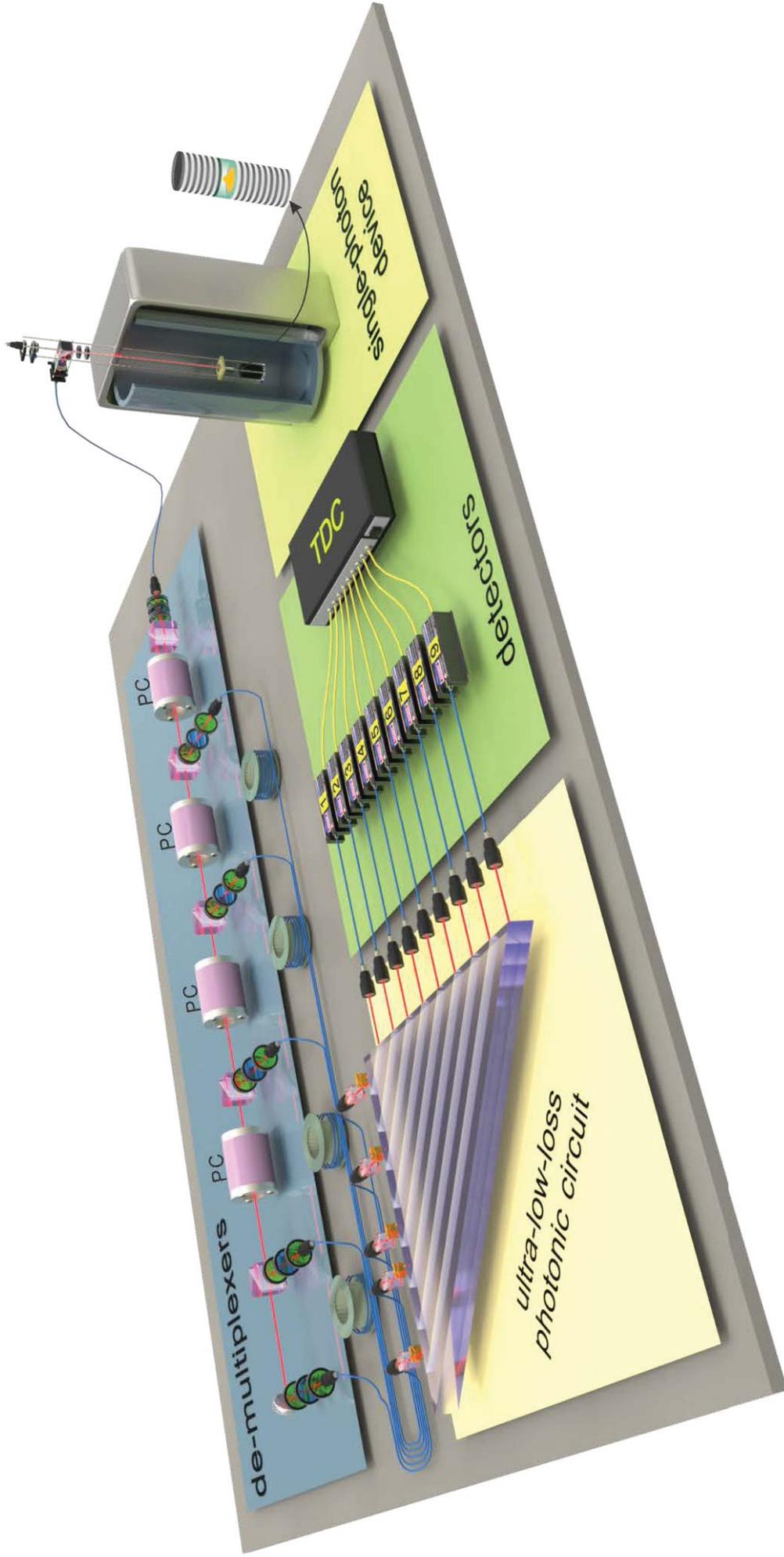

Figure 1

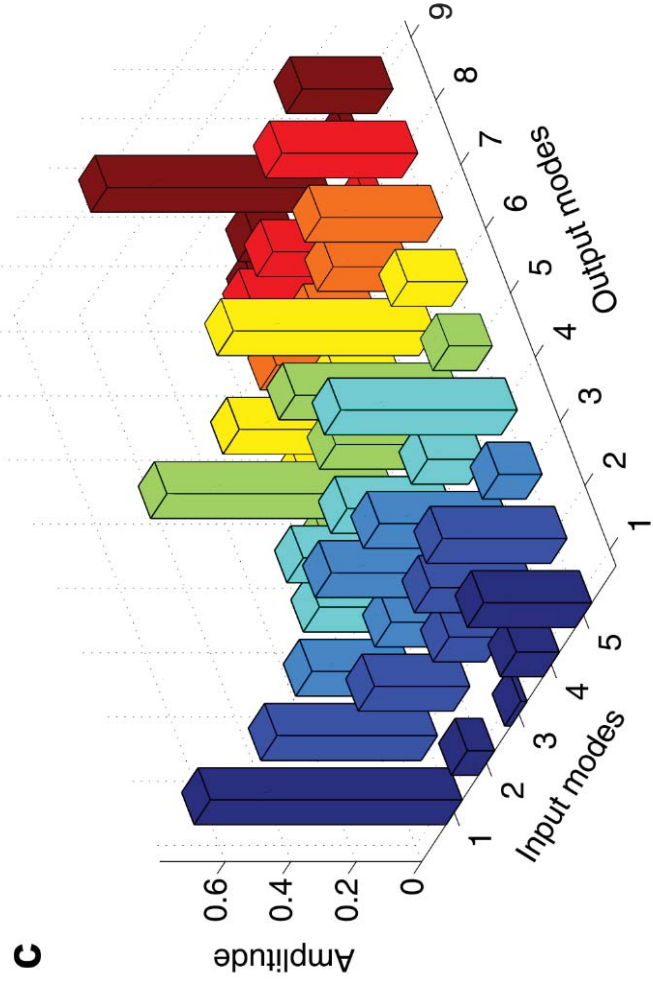
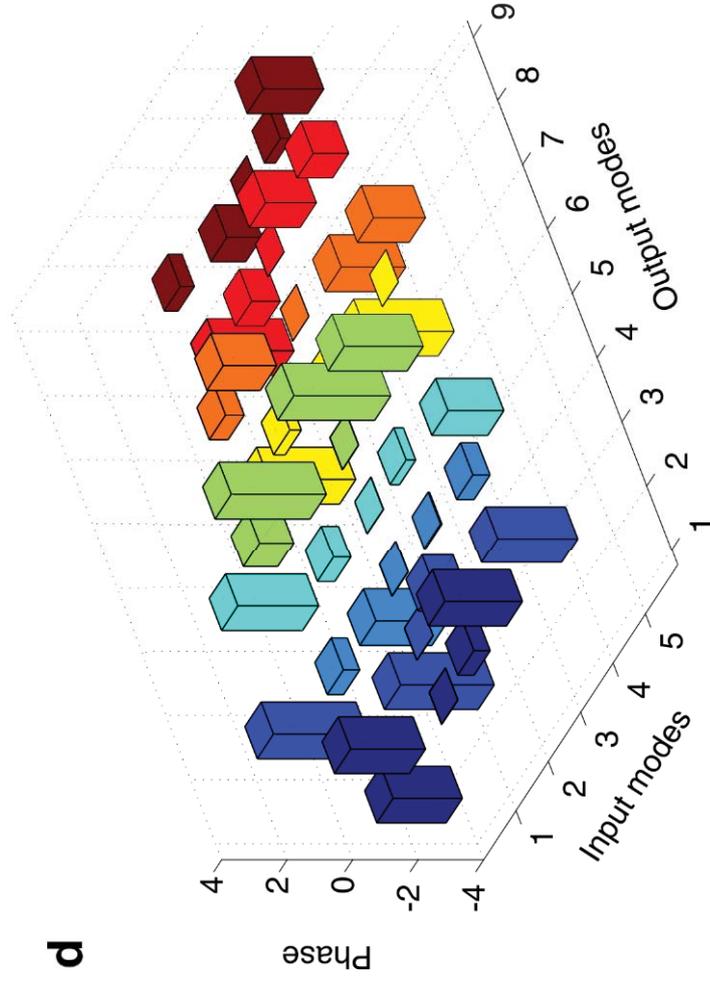
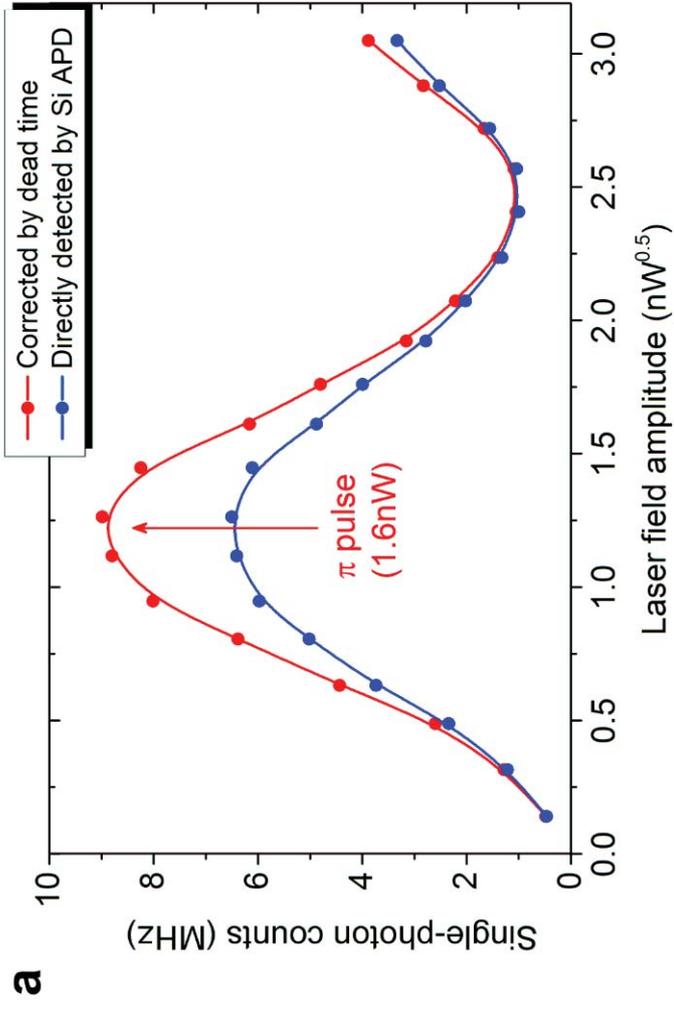
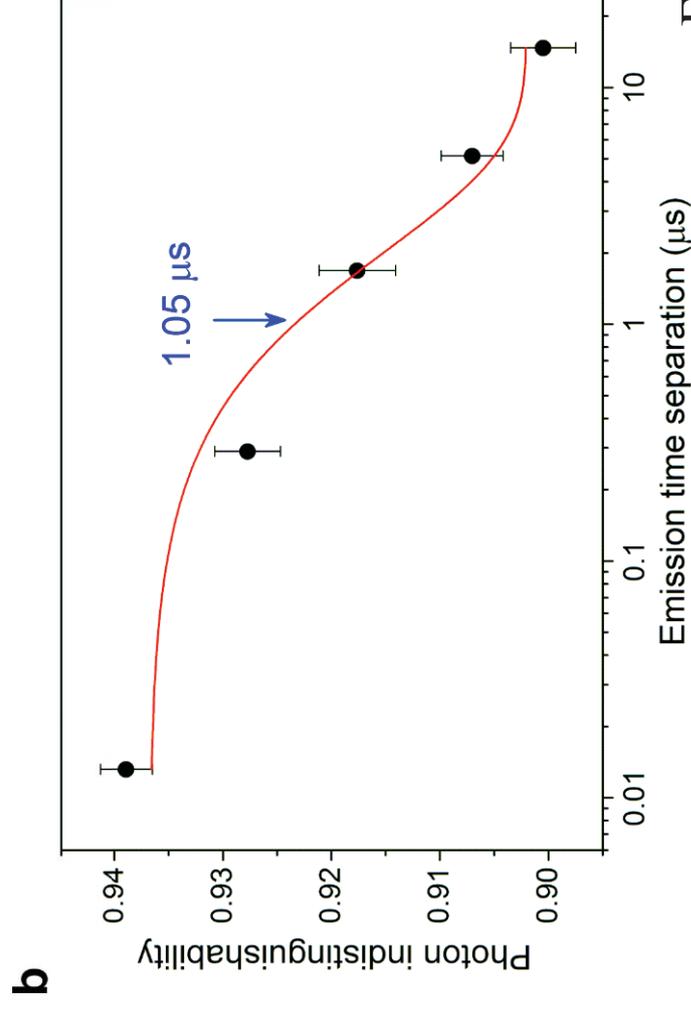

Figure 2

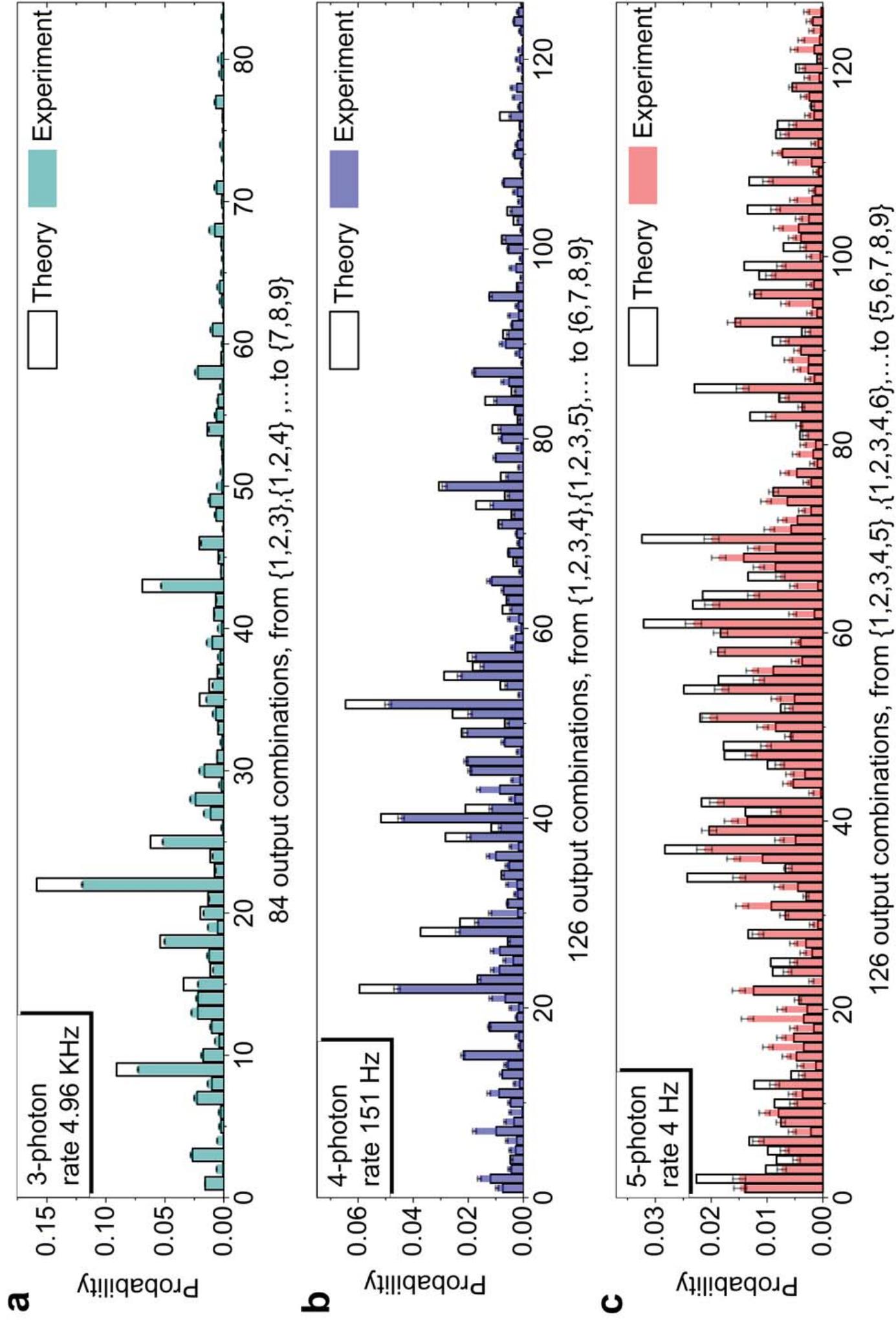

Figure 3

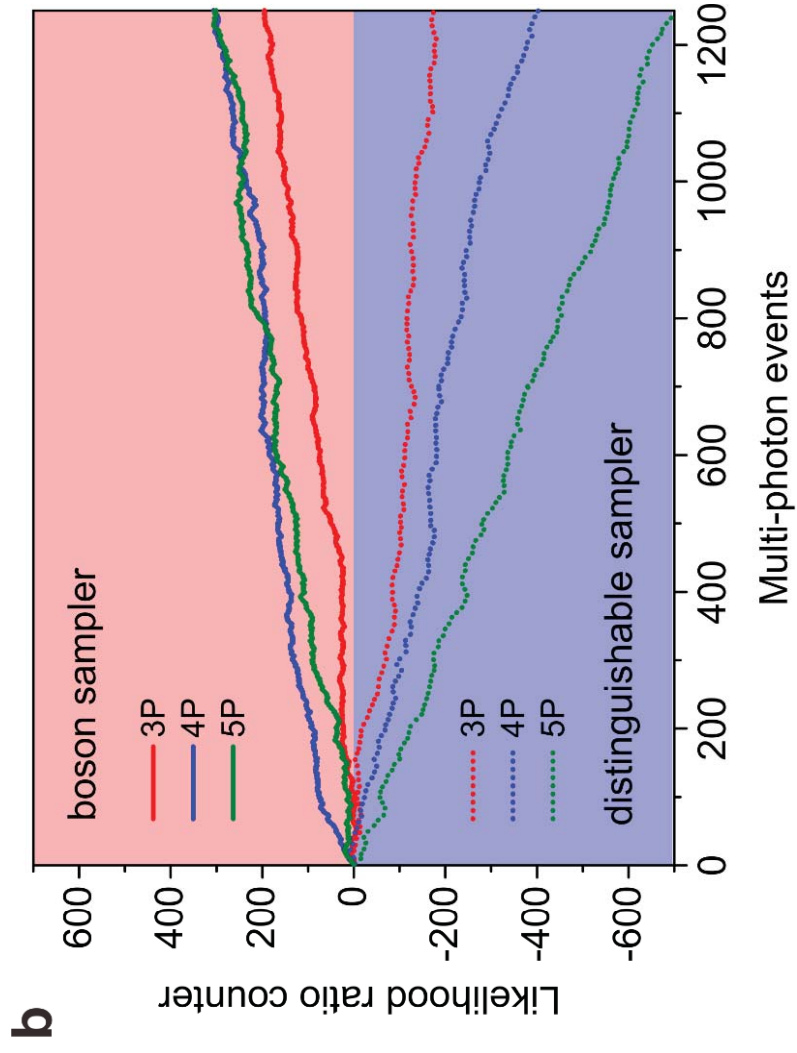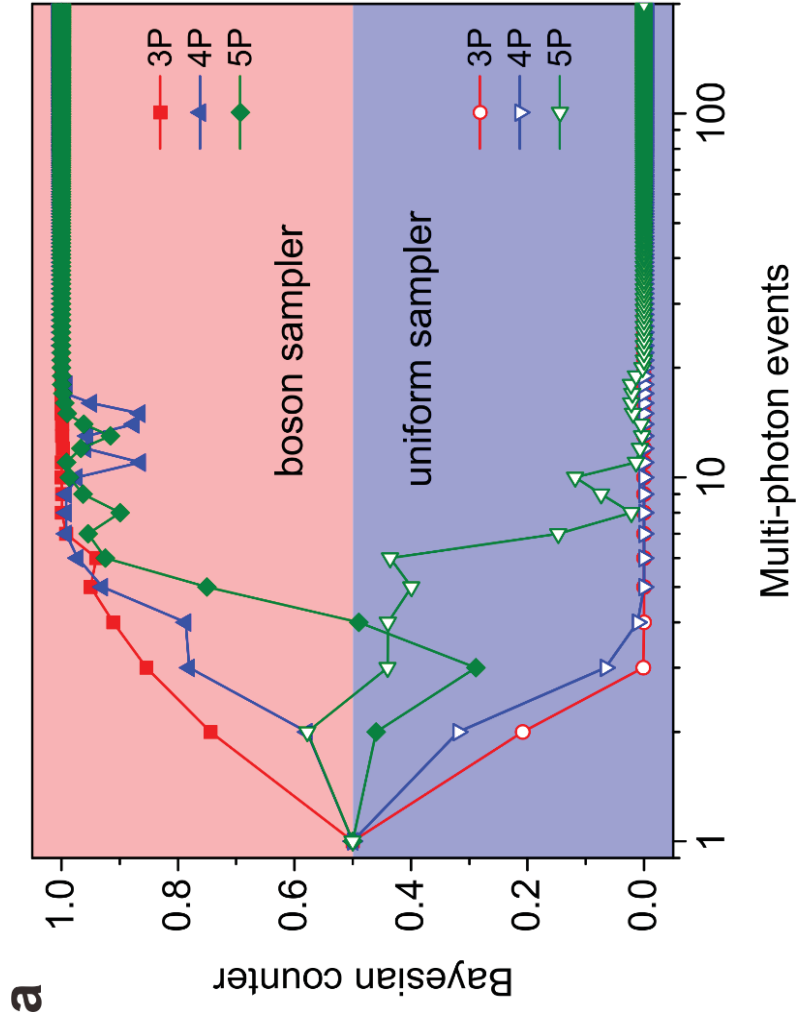

Figure 4

# Supplementary Information

Hui Wang[1,2*], Yu He[1,2*], Yu-Huai Li[1,2*], Zu-En Su[1,2], Bo Li[1,2], He-Liang Huang[1,2], Xing Ding[1,2], Ming-Cheng Chen[1,2], Chang Liu[1,2], Jian Qin[1,2], Jin-Peng Li[1,2], Yu-Ming He[1,2,3], Christian Schneider[3], Martin Kamp[3], Cheng-Zhi Peng[1,2], Sven Höfling[1,3,4], Chao-Yang Lu[1,2], and Jian-Wei Pan[1,2]

[1] Hefei National Laboratory for Physical Sciences at Microscale and Department of Modern Physics, University of Science and Technology of China, Hefei, Anhui, 230026, China

[2] CAS-Alibaba Quantum Computing Laboratory, CAS Centre for Excellence in Quantum Information and Quantum Physics, University of Science and Technology of China, China

[3] Technische Physik, Physikalisches Instität and Wilhelm Conrad Röntgen-Center for Complex Material Systems, Universitat Würzburg, Am Hubland, D-97074 Würzburg, Germany

[4] SUPA, School of Physics and Astronomy, University of St. Andrews, St. Andrews KY16 9SS, United Kingdom

* These authors contributed equally to this work

The supplementary information includes:

1. A summary on performance of boson-sampling experiments
2. Quantum dot sample details and experimental setup
3. Characterization of the single-photon source
4. De-multiplexing the single-photon sources into $N$ modes
5. Ultra-low-loss optical network fabrication and characterization
6. Validation: Bayesian method and Likelihood ratio test
7. An imagined race with ENIAC and TRADIC
8. Estimation of boson-sampling rate with larger number of photons

Supplementary table I and II

Supplementary figures 1, 2, 3, 4, 5, 6, 7.

# 1. A summary on performance of boson-sampling experiments

**Supplementary Table S1: Summary of the multi-photon boson-sampling experiments, and the key parameters relevant to the performance of boson-sampling.**

| Ref | Photon sources | Input photon state | photon indistinguishability | Single-photon system efficiency* | Network mode number | Network efficiency[#] | Detection efficiency | Final N-fold count rate (Hz) | Remarks |
|---|---|---|---|---|---|---|---|---|---|
| [1] | SPDC | $\|1\rangle\|1\rangle\|1\rangle$ | $$ | <0.01 | 6 | 0.64 | ~0.6 | 0.185 | Fiber[&&] |
| [2] | SPDC | $\|1\rangle\|1\rangle\|1\rangle$ | ~0.9 | <0.011 | 5 | 0.21 | ~0.65 | ~0.07 | |
| [3] | SPDC | $\|1\rangle\|1\rangle\|1\rangle$ | $$ | <0.011 (pair generation) | 6 | N/A | >0.5 | ~0.003 | |
| | SPDC | $\|2\rangle\|2\rangle$ | | <0.023 (pair generation) | 6 | N/A | >0.5 | ~0.001 | |
| [4] | SPDC | $\|1\rangle\|1\rangle\|1\rangle$ | 0.63 for different pairs | N/A | 5 | 0.3 | ~0.65 | N/A | |
| | SPDC | $\|1\rangle\|1\rangle\|1\rangle$ | | N/A | 9 | 0.05 | ~0.65 | N/A | |
| [5] | SPDC | $\|1\rangle\|1\rangle\|1\rangle$ | 0.99 for same pairs, 0.91 for different pairs | N/A | 21 | 0.3 | ~0.65 | N/A | |
| | SPDC | $\|\psi_{4\phi}\rangle$** | | N/A | 21 | 0.3 | ~0.65 | N/A | |
| | SPDC | $\|\psi_{5\phi}\rangle$** | | N/A | 21 | 0.3 | ~0.65 | N/A | |
| [6] | SPDC | $\|1\rangle\|1\rangle\|1\rangle$ | $$ | <0.023 | 13 | N/A | ~0.65 | N/A | |
| [7] | SPDC | $\|1\rangle\|1\rangle\|1\rangle$ | ~0.9 | <0.013 | 5 | <0.21 | ~0.65 | ~0.15 | |
| [8] | SPDC | $\|1\rangle\|1\rangle\|1\rangle$ | 0.95 for same pairs, ~0.7 for different pairs | ~0.014 | 13 | ~0.13 | ~0.65 | 0.008 | Scatter-shot |
| [9] | SPDC | $\|1\rangle\|1\rangle\|1\rangle$ | $$ | N/A | 6 | 0.58 | ~0.65 | N/A | ## |
| | SPDC | $\|3\rangle\|3\rangle$ | | N/A | | N/A | ~0.65 | N/A | |
| [10] | QD | $\|1\rangle\|1\rangle\|1\rangle$ | 0.5-0.7 | 0.14 | 6 | N/A | 0.32 | 0.21 | %% |
| This work | QD | $\|1\rangle\|1\rangle\|1\rangle$ | 0.923-0.939 | 0.337 | 9 | 0.99 | 0.32 | 4960 | |
| | | $\|1\rangle\|1\rangle\|1\rangle\|1\rangle$ | | | | | | 151 | |
| | | $\|1\rangle\|1\rangle\|1\rangle\|1\rangle\|1\rangle$ | | | | | | 4 | |

\* The "single-photon system efficiency" can be calculated by the rate of single photons arriving at the output of a single-mode fiber (or, input of multi-mode interferometer) divided by the repetition rate. It include the single-photon generation efficiency, heralding efficiency, collection efficiency, and all other loss in the optical channel.

# The "network efficiency" refers to the ratio between photons output from the interferometer and input into the it. It include both coupling efficiency and propagation efficiency.

$$ Data not provided in the manuscript. From the publications of the same groups, at optimal experimental condition (at low pump power, with narrowband (~3nm) filters, and good spatial mode-match), typical values of SPDC photon indistinguishabilities were ~0.98 for the same pair, and ~0.90 for independent pairs.

&& Polarizing fiber beam-splitters, works by evanescent coupling between multiple input fibers in close proximity. Note the network matrix is uncontrollable, random (fixed after fabrication) and not tunable.

\*\* $|\psi_{4\phi}\rangle = e^{i(\phi_1+\phi_2+\phi_3+\phi_4)}|1111\rangle + e^{2i(\phi_1+\phi_2)}|0022\rangle + e^{2i(\phi_3+\phi_4)}|2200\rangle$,

$|\psi_{5\phi}\rangle \approx |1122\rangle + |2211\rangle + |3300\rangle + |0033\rangle$. All the four modes were sent to a fully connected photonic circuit. Note that these photons cannot be treated as four or more *independent* single photons in a single run experiment. The way to simulating multiple single photons' evolution is as following: implement a series of experiments with different input conditions, calculate over all the experimental data, and reproduce the same output statistics as that of independent single-photons input. This method is not scalable.

## The multi-mode interferometric network demonstrated in this work is arbitrarily reprogrammable and universal for linear optics.

%% The photon indistinguishability in ref. [10] is limited to 0.5-0.7 due to the non-resonant optical excitation of the quantum dot. The indistinguishability was ~0.7 at low excitation power (but with low generation efficiency), and 0.5 at high excitation power. In addition, ref. [10] use beam splitters to passively de-multiplex the single-photon steams into different spatial modes, with an exponentially loss of efficiency.

N/A: Data not provided in the publications.

## 2. Quantum dot sample details and experimental setup

The quantum emitter used in our work is a self-assembled InAs/GaAs quantum dot which is grown by molecular beam epitaxy. To enhance photon extraction efficiency, the quantum dot is embedded in a λ-thick GaAs cavity and sandwiched between 25.5 and 15 λ/4-thick AlAs/GaAs mirror pairs forming the lower and upper distributed Bragg reflectors[11]. Micropillars with 2-μm diameter are defined through electron beam lithography and dry etching. The device is mounted on a three-dimensional piezo-electric 'slip-stick' positioner (Attocube), and cooled down to a temperature of 4-30 K inside an ultra-stable cryogenic-free bath cryostat (Attodry1000). We first characterize the device through photoluminescence measurements upon a 780-nm laser excitation. The temperature-dependent photoluminescence spectra are plotted in Supplementary Fig. 1. We perform time-resolved resonance fluorescence detection as a function of cavity-QD detuning. The extracted lifetime is plotted in Supplementary Fig. 2, from which we observe the lifetime of 60.6 ps at resonance, and we can deduce a Purcell factor of 7.63(23).

For pulsed *s*-shell resonant excitation on the quantum dot, we use a Ti:sapphire laser (Coherent Mira 900) at a central wavelength of 893.2 nm and with a pulse duration of ~3ps. To match the micropillar cavity linewidth, the excitation laser is further filtered with an etalon with a bandwidth of ~50 GHz.

As shown in Fig.1, the confocal microscope is operated in the cross-polarization configuration[12]. The excitation laser from the top arm transmits through a polarizing beam splitter (PBS), a half-wave plate (HWP) and a quarter-wave plate (QWP) and is focused on the quantum dot. The resonance fluorescence passes through the QWP, HWP and then is reflected by the PBS and collected from the side arm. By finely rotating the angles of the HWP and QWP, the scattering laser background can be suppressed by a factor exceeding $1 \times 10^7$. The ratio between the resonance fluorescence photons to the remaining laser scattering background is about 78:1. All optical elements in the optical path, except the optical window of the cryostat, are anti-reflection coated, in order to enhance the end-user single-photon efficiency (that is, the eventually obtained single photons out of the single-mode fiber).

## 3. Characterization of the single-photon source

The generated single photons have an eventual count rate on a silicon single-photon detector of 6.5 MHz without correcting the dead time, and 9 MHz after correcting the dead time. They are spectrally measured to be a single-frequency line with a full width at half maximum of 2.86(4) GHz (see Supplementary Fig. 3). These photons are directly used—without any spectral filtering—for the $g^2(\tau)$, two-photon interference, and boson-sampling measurements.

A key prerequisite for the generated photons to be useful for boson-sampling is that they should possess both high purity and indistinguishability[13]. Pure single-photon Fock states should have no multi-photon admixture. Supplementary Fig. 4a shows the data of second-order correlation measurement of the $\pi$ pulse-driven resonance fluorescence photons. At zero time delay, it shows a clear antibunching with a small multiphoton probability of 0.027. The non-classical Hong-Ou-Mandel interference in the boson-sampling multi-photon interferometry relies on a high degree of indistinguishability between the photons that emit with a large time separation[14]. We send the single photons into an unbalanced Mach-Zehnder interferometer so that the two photons are separated by a time delay from 13 ns to 14.7 μs. The two photons, with their polarization prepared either parallel or orthogonal, are overlapped on a beam splitter, and the two-photon output coincidence counts are measured. Supplementary Fig. 4b shows the time-delayed histograms of normalized two-photon counts for cross (black) and parallel (red) polarization at a time separation of 13 ns, where we observe a significant suppression of the counts at zero delay for parallel polarization and obtain a degree of indistinguishability of 0.939(3). We further measured at a time delay of 290 ns, 1.68 μs, 5.15 μs, and 14.7 μs, where the photon indistinguishability slightly drops to 0.928(3), 0.918(4), 0.907(3), and 0.900(3), respectively (see Fig. 2b in main text). The time scale of the decrease is fitted to be ~2.1 μs. In our boson-sampling measurement, the longest separation time of two de-multiplexed single photons is 80 pulse sequences, ~1.05 μs, where the photon indistinguishability remains 0.923.

## 4. De-multiplexing the single-photon sources into *N* modes

To perform multi-photon boson-sampling using a single-photon source, first we de-multiplex the pulsed train into *N* spatially separated modes. We note that our method

eliminates the need for growth and control of homogeneous quantum dots[15], which was known challenging due to the self-assemble process. The de-multiplexer consists of a Pockels cell (PC) and a PBS. Each PC (EKSMA Optics) contains two potassium titanyl phosphate (KTP) crystals. Under a half-wave voltage (~2000 V), the PC rotates horizontal ($H$) to vertical ($V$) polarization. When no electric field is applied, the photon polarization remains unchanged. We call the status of the PC as ON and OFF, when it is applied by 2000 V and 0 V, respectively. After careful compensation of the birefringence of the two KTP crystals, the extinction ratio of each PC is better than 100:1. The PBS behind the PC reflects the $V$ polarized photons to each separated spatial mode, and transmits the $H$ polarized photons to the next PC.

The KTP crystals are antireflection coated and have a high transmission ratio of 99.5%. The PCs are driven by high voltage pulses with an amplitude of ~2000V and a rise/fall time of ~8 ns, shorter than the period (~13 ns) between two single photons. The PCs are operated at a repetition frequency of 0.76 MHz, 1/100 of the 76-MHz repetition rate of the single photon pulse. The PC pulse sequence is shown in Supplementary Fig. 5. At the start point, all the single photons are prepared in the $H$ polarization. In each period of 1.3 μs, a voltage pulse with 264-ns duration is applied on the first PC (status ON), steering 20 single-photon pulses through the reflected arm. After that, the first PC is switched to OFF, and the second PC is switched ON for 264 ns. This continues, and eventually we obtain 5 spatially-separate single-photon pulses.

To synchronize the arrival time of these photons from different paths to overlap on the interferometer, we use single-mode fibers of various length (217 m, 163 m, 110 m, 56 m, and 2.7 m) to compensate their relative delay. The overall de-multiplexing efficiency for each channel is measured to be 84.5% on average, mainly due to the loss in coupling and propagation in the optical fiber coupling. Using PCs with higher repetition rate (the current PCs are limited to 1 MHz), the length of the compensation fibers can be reduced and thus increase the average efficiency to above 90%. At the output of the optical fibers, we use a combination of QWP and HWP to compensate the polarization rotation induced in the optical fibers. Further, the fibers are temperature stabilized to be within 0.5 degree to avoid environmental fluctuations. In addition to the coarse adjustment using fibers, before feeding into the interferometer, translation stages with sub-micrometer resolution and 25-mm travelling range are used for fine adjustment of the temporal delay to ensure a perfect overlap on the beam splitter

network.

In addition, we need to synchronize the single-photon pulses with the PCs driving signals and time-to-digital converter (TDC) for multi-photon events registration. To do so, a small fraction of the pulsed laser is split and converted into 76-MHz electric signal by a fast photodiode. The electric signal is fed into a field programmable gate array (FPGA) board which serves as a frequency divider with division factor of 100 to produce two channels of 760 KHz trigger signal, one of which is used as the trigger of the PCs and the other one is directly recorded by the TDC. Four arbitrary function generators (AFG) control the driving voltage pulse sequences for the PCs. The trigger signal passes through the four AFGs one by one, where the output delay of each AFG is finely tuned to match the single-photon pulses.

5. **Ultra-low-loss optical network fabrication and characterization**

The linear optical network, which implements a unitary transformation to the input state comprising $N$ single photons, is one of the key elements in the boson-sampling experiment. The linear optical network can be composed of beam splitters and phase shifters. To reliably obtain a fast multi-photon sampling rate for demonstrating "quantum supremacy", the most relevant criteria for linear optical networks are that they should be designed to implement a large, Haar random unitary matrix, robust (phase stable) to environmental fluctuations and have an ultra-high transmission rate. Other desirable features include miniaturized chip-size and universally programmable. It's known that the tradition networks consist of more than tens of bulk beam splitters are unstable and not scalable. A promising approach is to integrate all the elements on a tiny photonic chip which features a high density of integration. However, the overall system efficiencies of these photonic chips[2-9] are so far still limited up to ~30%, which significantly decreased the multi-photon count rates (see Supplementary Table 1).

In our work, we put forward a new circuit design that simultaneously combines the near-perfect stability, near-unity transmission rate, and randomness of the matrix. As shown in Supplementary Fig. 6, the triangular-shaped 9×9 mode interferometer consists of 36 beam splitters and 9 mirrors with a size of $37.8\text{mm} \times 37.8\text{mm} \times 4.2\text{mm}$, much smaller than the typical meter-size bulk optics but larger than the millimeter-size silica waveguide photonic chips[2-9]. The fabrication process is as following. A series of trapezoid-shaped fused quartz plates are cut and finely polished. The top surfaces of

each trapezoids (L1-L8) are optically coated with polarization-dependent beam-splitting ratios. For the *V* and *H* polarization, the transmission to reflection ratios are 0.42:0.58 and 0.9:0.1, respectively. Next, we bond these separate plates together via intermolecular force. Finally, the outer surface L9 is total-reflection coated, and L10 and L11 are polished again and antireflection coated to further improve the transmission efficiency. These plates are designed with dimensional tolerance below 5 μm and angle variation below 24 μrad to ensure a good parallelism between the neighboring plates.

We experimentally calibrate the key performance of this 9×9 mode interferometer, including its transmission efficiency, spatial-mode matching, stability and adjustability. Thanks to the antireflection coating, the overall transmission efficiency (from input to output) is measured to be above 99%. By Mach-Zehnder-type coherence measurements, the spatial-mode overlap is determined to exceed 98% using free-space photodetection, and be better than 99.9% with single-mode fiber coupling (at the cost of ~5% efficiency loss). The compact interferometer is housed on a temperature-stabilized baseplate, and remains stable at least for weeks. An example is that we perform two five-photon boson sampling tests with a time separation of five days, and the two results show a similarity of 99.5% (see Supplementary Fig. 7). This linear optical network is not fully universal as in Ref. (19), however, its beam splitting ratio can be continuously adjusted by changing the input photons' polarization. Therefore, many different random matrices can be obtained. Such a design can be upgraded to a loss-tolerant, square-shape design recently proposed by Walmsley's group[16], further miniaturized, electrically tuned, and scaled to reasonably larger dimensions, and may be suitable for the near-term goal of demonstrating "quantum supremacy".

## 6. Validation

### a) Bayesian method

We use Bayesian analysis[17] method to discriminate a boson-sampler from other hypotheses, such as uniform sampler and distinguishable sampler. Let $Q$ be Boson sampling, and $R$ be an alternative hypothesis. For each measured event $k$, let $q_k$ and $r_k$ be the probability associated with hypothesis $Q$ and $R$ respectively. After $N_{events}$ events, according to the Bayes' theorem, we can get

$$\frac{P(Q|N_{events})}{P(R|N_{events})} = \prod_{x=1}^{N_{events}} (\frac{q_x}{r_x}) = \chi$$

Then the probability of this $N_{events}$ events assigned to boson sampler is

$$P(Q|N_{events}) = \frac{\chi}{\chi+1}$$

If the $N_{events}$ events are from a genuine boson sampler, then $P(Q|N_{events})$ tends to 1, while $P(R|N_{events})$ tends to 0.

**b) Likelihood ratio test**

We apply likelihood ratio test[6,18] to exclude the distinguishable sampler in the main text. Let $p_k^{ind}$ and $q_k^{dis}$ denote the probabilities associated with indistinguishable and distinguishable photons for the observed event $k$, respectively. For each measured event $k$, we calculate the estimator $L_k = p_k^{ind}/q_k^{dis}$, and a counter $C$ initialized to 0 and updates as following:

$$C := \begin{cases} C, & a_1 < L_k < 1/a_1 \\ C+1, & 1/a_1 \leq L_k < a_2 \\ C+2, & L_k \geq a_2 \\ C-1, & 1/a_2 \leq L_k < a_1 \\ C-2, & L_k \leq 1/a_2 \end{cases}$$

In our test, we set $a_1 = 0.85$ and $a_2 = 1.8$. After $N$ events, if $D > 0$, the test decides the data are from indistinguishable bosons, otherwise the data are from distinguishable bosons.

**7. An imagined race with ENIAC and TRADIC**

It would be interesting to imagine a race between our small-size quantum boson-sampling machines with the earliest classical computers, following the rules defined in Ref. 1 and 3. Here we directly quote from Ref. 1: *"Alice, who only possesses classical resources, and Bob, who in addition possesses quantum resources. They are given some physical operation, described by an evolution operator, U, and agree on a specific n-boson input configuration. Alice calculates an output sample distribution with a classical computer; Bob either builds or programs an existing linear photonic network, sending n single photons through it and obtaining his sample by measuring the output*

*distribution. The race ends when both return samples from the distribution: The winner is whoever returns a sample fastest"*. For classical computers, it is assumed that obtaining one sample requires the calculation of one permanent.

To avoid misinterpretations, we stress again that a quantum boson-sampling machine can't calculate the permanent efficiently, as boson-sampling is a sampling problem, not a decision problem. It relates to calculating the permanent but it is itself not a #P-complete problem.

In human history, the first electronic computer—ENIAC—performs 5000 additions or 357 multiplications per second, whereas the first transistorized computer—TRADIC—performs 62500 additions or 3333 multiplications per second. The classical benchmarking algorithm for calculating the permanent of an $n \times n$ matrix $X$ is the well-known Ryser's formula[19]: $\mathrm{per}(X) = \sum_{\varepsilon_j \in \{0,1\}} (-1)^{n-\varepsilon_1-\cdots-\varepsilon_n} \prod_{1 \leq i \leq n} \sum_{1 \leq j \leq n} \varepsilon_j x_{ij}$, which requires $(2^n-1)(n-1)$ multiplication operations and $(2^n-2)(n+1)$ addition operations, respectively[20]. Using the published gate time of the ENIAC[21] and TRADIC[22], we can estimate the run time using the classical algorithm to obtain one permanent. The time required to obtain one sample using the quantum machine and calculate the submatrix using the simulated classical computer are listed in Supplementary Table S2. In this sense, for a comparison, the 3-boson sampling machine is 220 times faster than ENIAC, and 23 times faster than TRADIC. The 4-boson sampling macine is 21 times faster than ENIAC, and 2.2 times faster than TRADIC. The 5-boson sampling machine is 1.5 times faster than ENIAC.

Note, however, that the above comparison doesn't consider the experimental error of the quantum machine, which would seem unfair. In our experiment, the upper bounds of additive error $\varepsilon$ of the obtained permanents, defined as $\left|\sqrt{p_i} - \sqrt{q_i}\right| < \varepsilon$ for all possible $\boldsymbol{p_i}$ and $\boldsymbol{q_i}$, are calculated to be 0.052, 0.065 and 0.041 for the 3-, 4- and 5-boson sampling, respectively. Taking the experimental error of the quantum device into account, the task of the classical machine can be accordingly changed to calculating *approximate* permanents. For this purpose, we use Gurvits's approximation algorithm[23], which takes $O(n^2/\varepsilon^2)$ time. In principle, for a sufficiently large system, approximate

permanents would be faster to calculate than perfect permanets, although still being intractable. However, for the current small-size system, approximate permanent takes even longer time, as listed in Supplementary Table II. We can see that in both cases, our quantum machines can obtain a sampling faster than the ENIAC and TRADIC.

Supplementary Table II. Required run time to precisely (using Ryser algorithm) or approximately (using Gurvits algorithm) calculate one permanent by classical computers or to obtain one sample by our quantum machines (unit: milliseconds).

|  | 3-boson sampling ($3\times3$ submatrix) | 4-boson sampling ($4\times4$ submatrix) | 5-boson sampling ($5\times5$ submatrix) |
| --- | --- | --- | --- |
| ENIAC[a] (Ryser) | 44.0 | 140.0 | 383.3 |
| ENIAC[a] (Gurvits) | 3773.4 | 3409.3 | 11306.1 |
| TRADIC[b] (Ryser) | 4.6 | 14.6 | 40.1 |
| TRADIC[b] (Gurvits) | 386.1 | 344.6 | 1130.4 |
| Our multi-photon interferometry | 0.2 | 6.6 | 248.8 |

[a]The ENIAC could do 5000 additions or 357 multiplications per second[21].
[b]The TRADIC could do 62500 additions or 3333 multiplications per second[22].

8. **Estimation of boson-sampling rate with larger number of photons**

The count rate (CR) of $n$-photon boson sampling can be expressed as

$$CR(n) = \frac{R_{pump}}{n}(\eta_{QD}\eta_{de}\eta_{C}\eta_{det})^n S$$

where $R_{pump}$ is the pumping repetition rate of the single-photon source, $\eta_{QD}$ is the single-photon source end-user brightness, $\eta_{de}$ is the de-multiplexing efficiency for each channel, $\eta_C$ is the average efficiency of the photonic circuit including the output coupling efficiency, and $\eta_{det}$ is the efficiency of the detectors. $S$ is the ratio of no-collision events to all possible output combinations. We note that $S$ value is relate to the specific unitary matrices, which can be estimated by $S = \binom{m}{n}\bigg/\binom{m+n-1}{n}$.

In this experiment, $R_{pump} = 76$ MHz, $\eta_{QD} = 0.338$, $\eta_{de} = 0.845$, $\eta_C = 0.905$, and $\eta_{det} = 0.32$. By antireflection coating on the optical window of the cryostat, we can straightforwardly improve the $\eta_{QD}$ to ~0.37. We can upgrade the silicon photon detectors with commercially available superconducting nanowire detectors with reported efficiency of ~0.95. Using photonic circuit with $4n$ spatial modes, we can reach coincident count rates of 144/h, 27/h, 5/h and 1/h for future 12-, 13-, 14- and 15-boson sampling, respectively.

On top of the above mentioned technical improvements which can be readily done, the single-photon source brightness can in principle be doubled, which can be achieved by removing the cross-polarization in the confocal setup for resonant optical excitation. The challenge is to cleanly extinguish the pumping laser background without the use of polarization filtering, which can be done by side excitation of the quantum dot. This can in principle improve the $\eta_{QD}$ to be ~0.74, and the 20-boson sampling rate can reach ~130/h.

**Supplementary Figure captions:**

**Supplementary Fig. 1** 2D intensity plot of temperature-dependent micro-photoluminescence spectra. The excitation cw laser is at 780 nm wavelength and the power is ~6 nW. The photoluminescence intensity reaches a plateau at a temperature range of 4–10 K.

**Supplementary Fig. 2** Measured pulsed resonance fluorescence (RF) lifetime as a function of QD-microcavity detuning by varying the temperature. The time-resolved data are measured using a superconducting nanowire single-photon detector with a time resolution of ~63 ps.

**Supplementary Fig. 3** A high-resolution resonance fluorescence spectrum obtained using a home-built Fabry-Perot scanning cavity with a finesse of 170, a linewidth of 220 MHz (full width at half maximum), a free spectral range of 37.4 GHz, and a transmission rate of 61%. The red line was a fit using a Voigt profile.

**Supplementary Fig. 4** Characterization of single-photon purity using intensity-correlation measurement (**a**) and indistinguishability using two-photon Hong-Ou-Mandel interference (**b**). The *y* axis are normalized coincidence rate and the *x* axis is the time delay between the two detectors. The two photons are prepared in parallel and perpendicular polarization in the red and blue curve of the lower panel, respectively.

**Supplementary Fig. 5** The voltage driving pulse sequence for the Pockels cell and the illustration of the active de-plexing single photon pulses.

**Supplementary Fig. 6 a**. Building a 9×9 mode ultra-low-loss photonic circuit from individual trapezoids with a bottom-up approach. L1, L2 … and L8 label the top surface of each trapezoid, which can be independently coated with polarization-dependent beam-splitting ratios. L9, L10, and L11 are the outer surface of the whole setup. Each trapezoid is finely polished to ensure a dimensional tolerance of less than 5 μm. We choose arbitrary five input modes (labelled by red solid circles) from the nine modes. **b.** The equivalent photonic circuit.

**Supplementary Fig. 7** A comparison between two 5-photon boson-sampling

experiments performed by a time separated by five days.

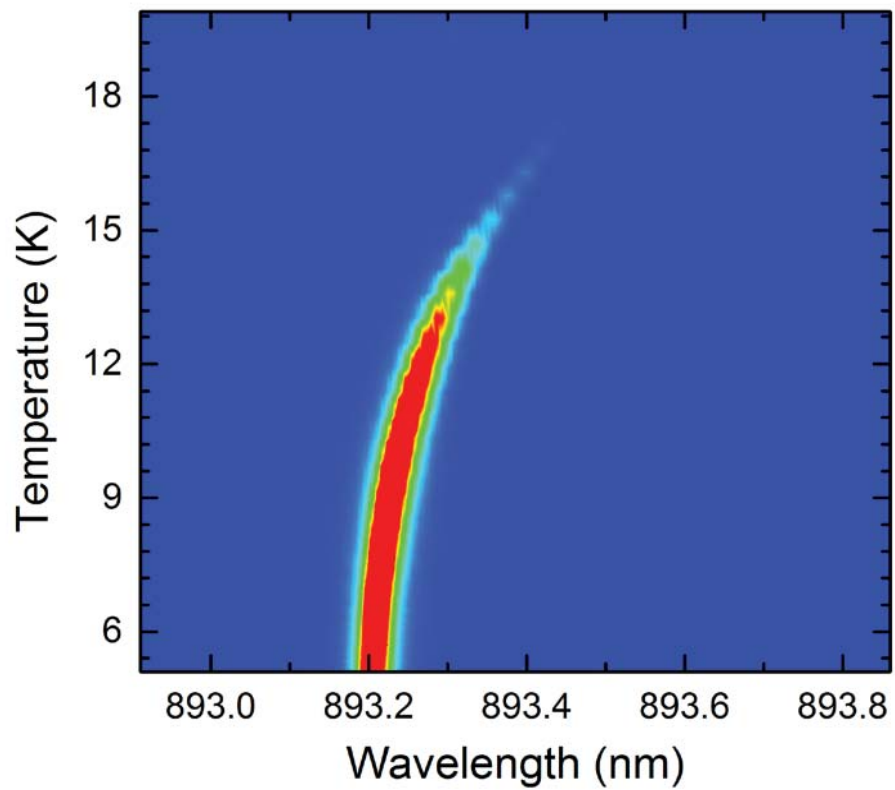

Supplementary Fig. 1

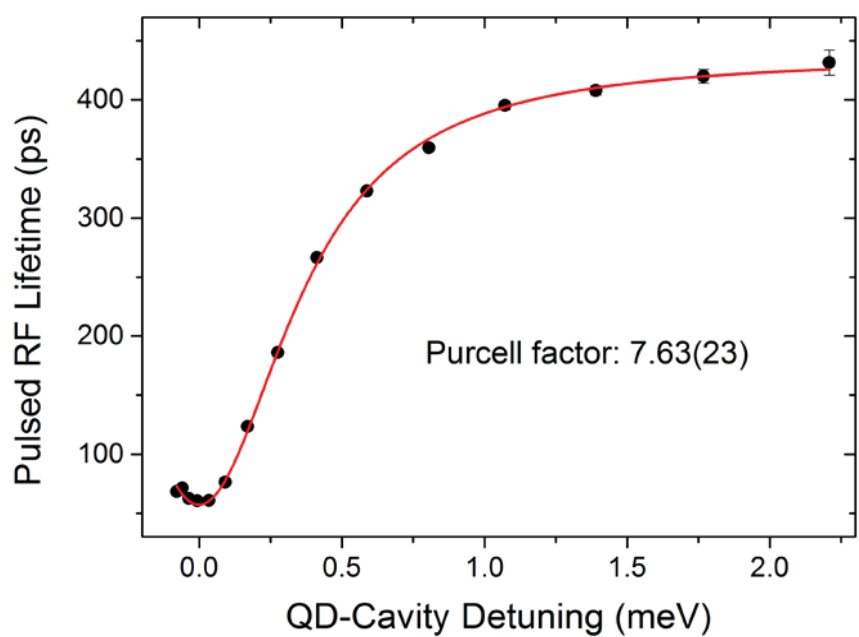

Supplementary Fig. 2

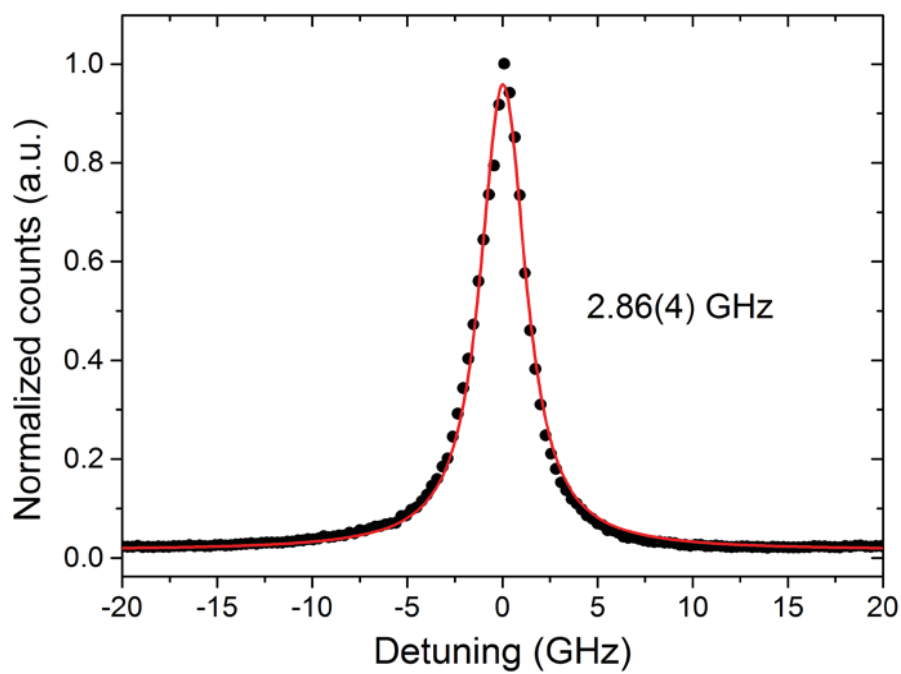

Supplementary Fig. 3

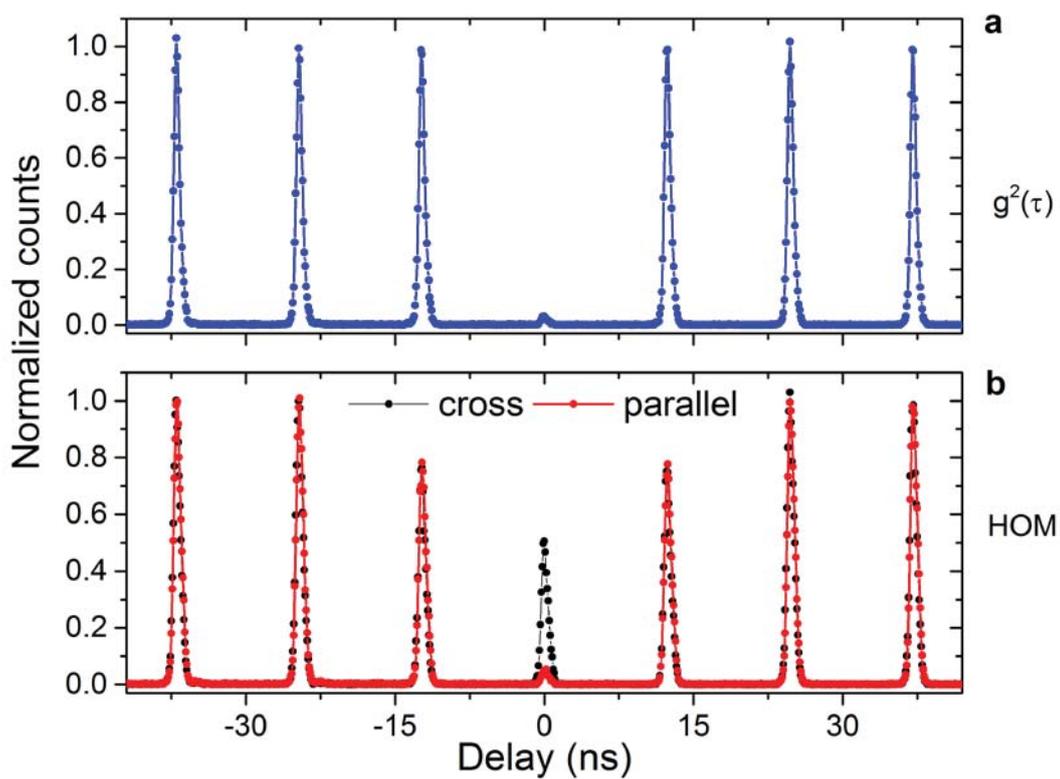

Supplementary Fig. 4

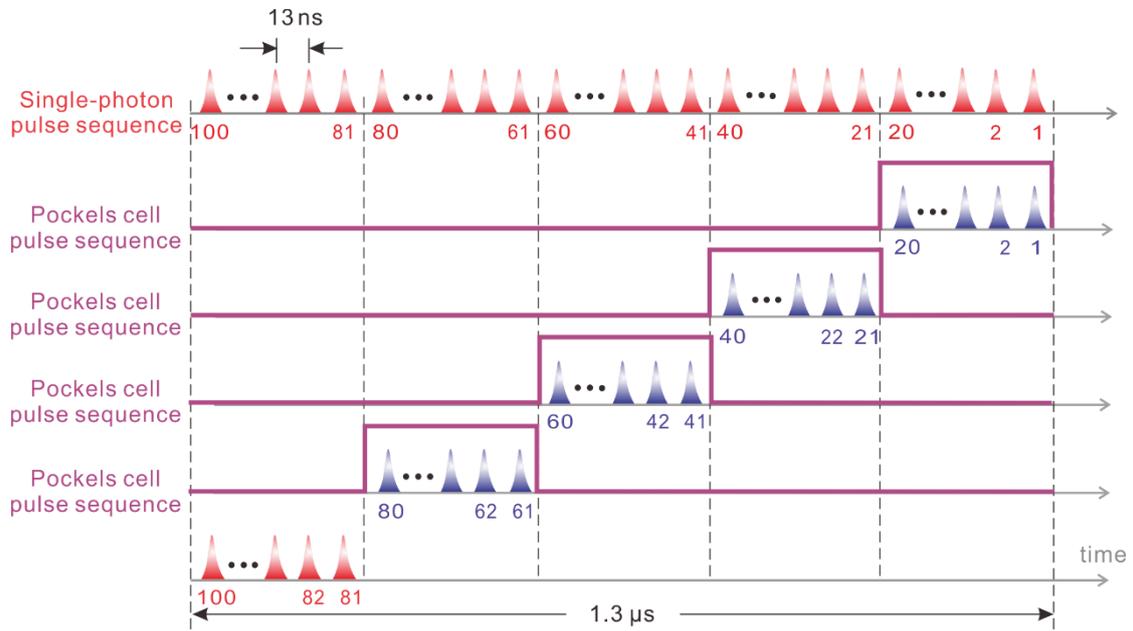

Supplementary Fig. 5

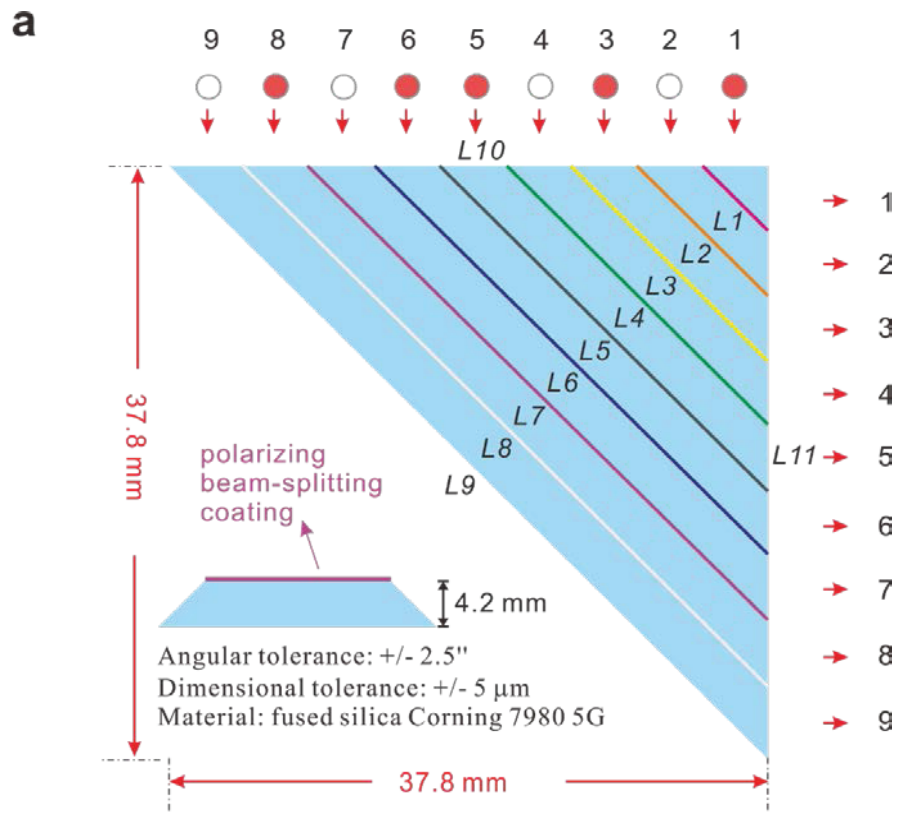

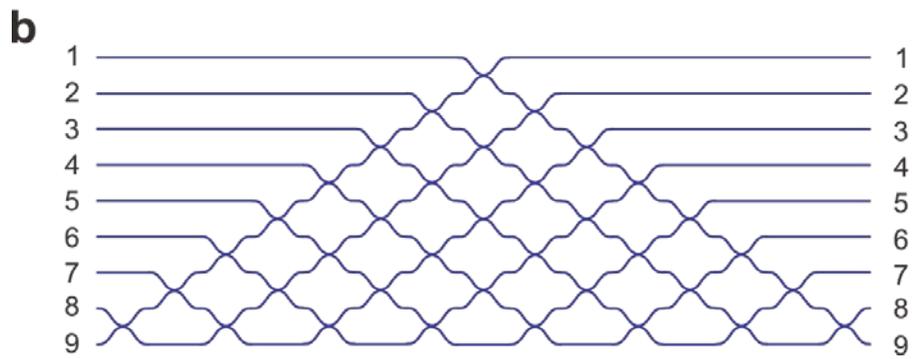

Supplementary Fig. 6

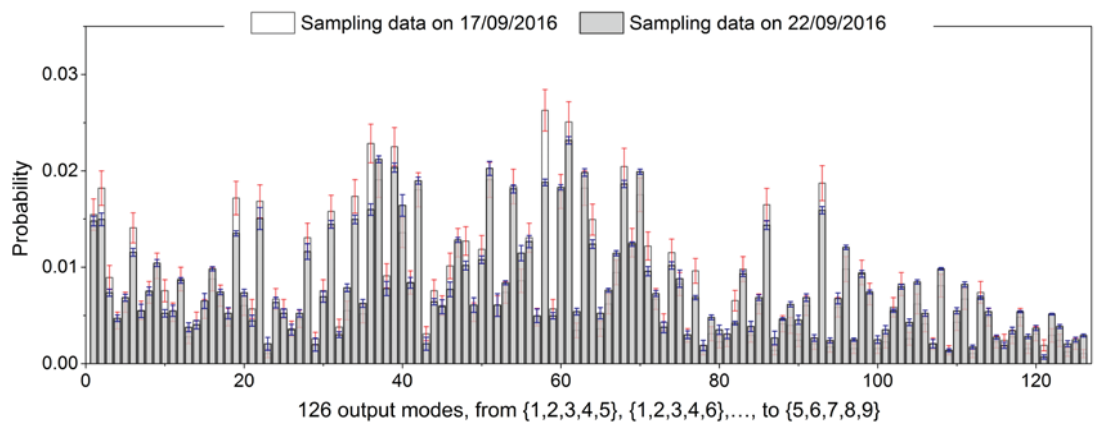

Supplementary Fig. 7